# Pressure-induced superconductivity in single crystal $CaFe_2As_2$

Tuson Park[1,2], Eunsung Park[2], Hanoh Lee[1], T. Klimczuk[1,3], E D Bauer[1], F Ronning[1], and J D Thompson[1]

[1] Los Alamos National Laboratory, Los Alamos, NM 87545 USA

[2] Department of Physics, Sungkyunkwan University, Suwon 440-746, Korea

[3] Faculty of Applied Physics and Mathematics, Gdansk University of Technology, Narutowicza 11/12, 80-952 Gdansk, Poland

**We report pressure-induced superconductivity in a single crystal of $CaFe_2As_2$. At atmospheric pressure, this material is antiferromagnetic below 170 K but under an applied pressure of 0.69 GPa becomes superconducting, with a transition temperature $T_c$ exceeding 10 K. The rate of $T_c$ suppression with applied magnetic field is -0.7 K/T, giving an extrapolated zero-temperature upper critical field of 10-14T.**

The tetragonal $ThCr_2Si_2$ crystal structure supports a wide range of elemental combinations, with the Th site alone being occupied by all the rare earths, the light actinides, divalent alkaline earths Ca, Sr and Ba and others. Recently, the $AFe_2As_2$ (A= Ca, Sr and Ba) materials that form in this structure [1] have received renewed attention [2-8] due to their similarity to another family of compounds ROFeAs (where R is a rare earth) that crystallizes in the tetragonal ZrCuSiAs structure type [9-14]. Both systems contain structural layers of FeAs, undergo a tetragonal to orthorhombic transition below room temperature and support antiferromagnetism in the orthorhombic structure. Non-isoelectronic chemical substitution in the out-of-plane layer, for example, substituting K for Ba in $BaFe_2As_2$ [7] or F for O in LaOFeAs [8], introduces additional charge carriers, suppresses the structural and magnetic transitions, and induces superconductivity at temperatures in the tens of Kelvins. Of these various materials, none with FeAs layers is superconducting without electronic doping, though the nominally isoelectronic nickel phosphides, $BaNi_2P_2$ [15] and LaONiP [16], with NiP layers do superconduct at relatively low (≤ 5K) temperatures. The unit cell volumes of the P-containing compounds are smaller than their As counterparts, and this suggests that an effective chemical pressure produced by replacing As with the smaller P atom may play a significant role. Thus far, no undoped, non-superconducting member of these families has been explored as a function of physically applied pressure; however, pressure does initially increase the superconducting transition temperature of F-doped LaOFeAs [17], lightly doped SmOFeAs [18] and oxygen-deficient LaOFeP [19]. These observations raise the question of whether it might be possible to induce superconductivity by applying sufficiently high pressures. Within the $AFe_2As_2$ family, the unit cell volume decreases monotonically in the sequence A = Ba, Sr, Ca.[4] Though none of these superconducts without chemical doping, the smallest volume member $CaFe_2As_2$ is a plausible candidate for initial exploration. In the following, we show that superconductivity does emerge in $CaFe_2As_2$ with only modest applied pressure.

Single crystals of $CaFe_2As_2$ were grown from a Sn flux and characterized as described in ref. [6]. Thin plate-like crystals form in the $ThCr_2Si_2$ structure with lattice parameters a = 3.887(4) Å and c = 11.76(2) Å. Specific heat, electrical resistivity, Hall coefficient and magnetic susceptibility reveal a first order phase transition at 171 K, below which the structure is orthorhombic and, in analogy with the Ba and Sr members of this family [4, 20, 21], the system is antiferromagnetic. Four-probe electrical resistance measurements were made on a single crystal, with current flow in the tetragonal plane. The crystal was mounted inside a Teflon cup within a hybrid BeCu/NiCrAl clamp-type pressure cell. The cup, filled with silicon oil, also contained a small piece of Pb whose known pressure-dependent superconducting transition [22] enabled a determination of the pressure within the cell.

Figure 1 compares the temperature dependence of the electrical resistivity at atmospheric pressure and at 0.69 GPa. In contrast to the obvious signature for the first order transition near 170 K at atmospheric pressure, the resistance at 0.69 GPa decreases smoothly and near 13 K begins to drop by over three orders of magnitude to an immeasurably small value below 8 K. In the absence of a signature for the first order transition at this pressure, we assume that superconductivity has developed in the

ThCr$_2$Si$_2$ structure, but this needs to be confirmed. Nevertheless, such a large change in properties with such a small applied pressure suggests that CaFe$_2$As$_2$ is very near an instability.

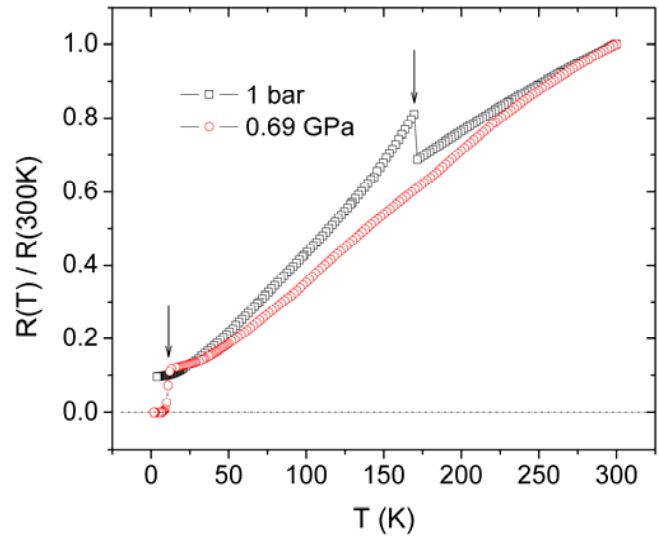

Figure 1 Temperature dependence of the normalized resistance of CaFe$_2$As$_2$. Resistance divided by its room-temperature value is plotted against temperature for 1 bar (squares) and 0.69 GPa (circles). For reference, the room temperature resistivity is approximately 0.22 mΩcm at atmospheric pressure. Arrows indicate magnetic and superconducting transitions for 1 bar and 0.69 GPa, respectively.

An expanded view of the resistive transition is plotted in figure 2(a) along with other curves measured in an applied field. The field-dependent evolution of the transition is consistent with it being one to a superconducting state. From the midpoint of the transition, we determine the field-temperature phase diagram shown in figure 2(b). For fields greater than ~ 1T, the linear rate of decrease in the superconducting transition temperature is -0.7 K/T (or equivalently $dH_{c2}/dT$= -1.4 T/K). An extrapolation of this slope to H=0 gives $T_c$ (H=0) = 10 K. However, as seen in this panel, the zero-field resistive midpoint transition temperature is higher, near 10.8 K. A high temperature tail in $H_{c2}(T)$ could be due to flux-flow induced by a measuring current that is comparable to the critical current density at these fields and temperatures; however, this seems to be an unlikely explanation in the present case because the measuring current density was small, about $10^{-4}$ A/cm$^2$. An alternative interpretation is that the tail is a manifestation of multi-band superconductivity [23, 24], a possibility we cannot rule out, and that would be consistent with Hall effect and thermoelectric power measurements [25] on crystalline CaFe$_2$As$_2$. In this scenario, $H_{c2}(T)$ should increase nearly linearly to very low temperatures and would give $H_{c2}(0) \approx 13.9$ T for CaFe$_2$As$_2$. For comparison, assuming CeFe$_2$As$_2$ is a single-band, dirty, type-II superconductor [26], we estimate $H_{c2}(0) = 0.7 T_c dH_{c2}/dT \approx 10$ T, which would require $H_{c2}(T)$ to decrease rather unrealistically by only about 10% between T=0 and ~ 0.3$T_c$.

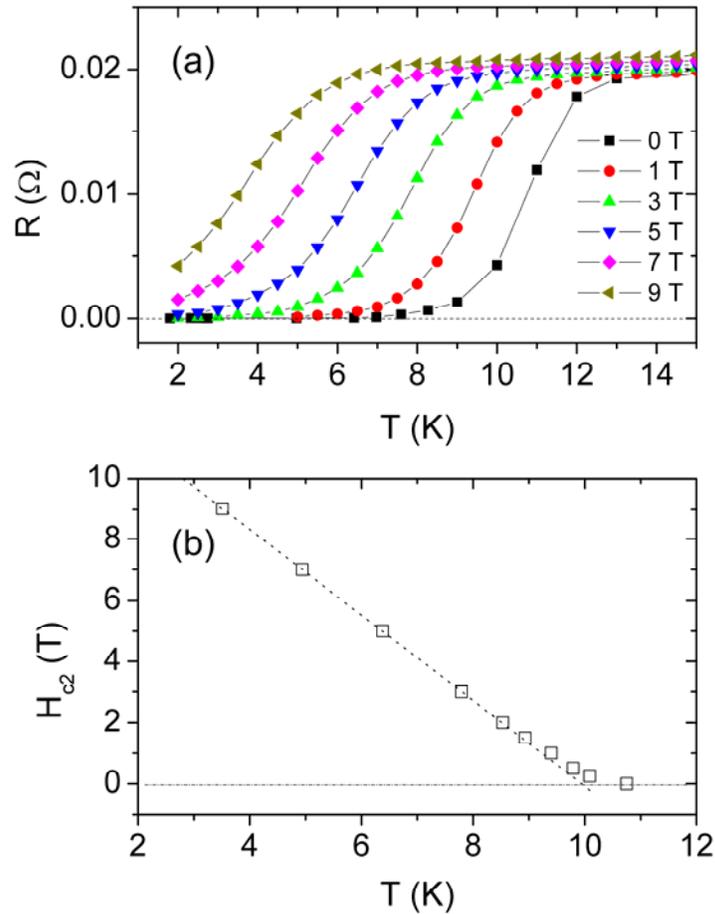

Figure 2 (a) Magnetic field dependence of the superconducting transition of CaFe$_2$As$_2$ at 0.69 GPa. (b) Temperature dependence of the upper critical field $H_{c2}$ at 0.69 GPa. The dotted line is a least-squares fit to a linear-T dependence with $H_{c2}(0) = 13.9$ T and $dH_{c2}/dT = -1.4$ T/K.

Na-substitution for Ca also induces superconductivity near 20 K in polycrystalline material with nominal composition $Ca_{.5}Na_{.5}Fe_2As_2$.[25] Like $CaFe_2As_2$ under pressure, the resistivity of this superconductor also lacks a signature for a first order phase transition and its unit cell volume is reduced. Which of these effects, hole doping or reduced cell volume, is more important for superconductivity and suppression of the first order phase transition requires further exploration as does their possible interplay. Considering larger cell volumes of the A= Sr, Ba compounds and the simple idea of chemical pressure, we might expect to find pressure-induced superconductivity in these systems near 15 GPa (Ba) and 8 GPa (Sr), assuming a calculated bulk modulus of 98 GPa [27]. Divalent Eu also forms in the $ThCr_2Si_2$ structure and band structure calculations on $EuFe_2As_2$ suggest that the well localized 4f electron of Eu is coupled only weakly to the FeAs sublattice [28], which appears to be the primary structural unit in these superconductors. With a cell volume intermediate to that of $SrFe_2As_2$ and $CaFe_2As_2$, divalent $EuFe_2As_2$ also might become superconducting at a pressure (~ 6 GPa) much lower than that estimated (50 GPa) to induce a change in the valence of Eu. [28]

In summary, $CaFe_2As_2$ is, to our knowledge, the first of the FeAs-based $ThCr_2Si_2$ structure-type materials to exhibit pressure-induced superconductivity. Its superconducting transition temperature at 0.69 GPa is relatively high, approaching 25-50% of the highest $T_c$s reported for hole doped $AFe_2As_2$ materials. With these observations, pressure offers a new route to superconductivity in these and possibly the related ZrCuSiAs materials without the need to introduce extrinsic chemical disorder.

Work at Los Alamos was performed under the auspices of the US Department of Energy, Office of Science and supported by the Los Alamos Laboratory Directed Research and Development program.